\documentclass[aps,prl,twocolumn,showpacs]{revtex4}

\newcommand{\Jeff}{J_{\mbox{\footnotesize eff}}}
\newcommand{\HBH}{H_{\mbox{\footnotesize BH}}}

\usepackage{graphicx}

\begin{document}

\title{Controlled generation of coherent matter-currents using a 
periodic driving field}

\author{C.E.~Creffield and F.~Sols}
\affiliation{Dpto de F\'isica de Materiales, Universidad
Complutense de Madrid, E-28040, Madrid, Spain}

\date{\today}


\begin{abstract}
We study the effect of a strong, oscillating driving field on the 
dynamics of ultracold bosons held in an optical lattice.
Modeling the system as a Bose-Hubbard model, we show
how the driving field can be used to produce and maintain a coherent
atomic current by controlling the phase of the intersite 
tunneling processes. We investigate both the stroboscopic
and time-averaged behavior using Floquet theory,
and demonstrate that this procedure provides a stable and 
precise method of controlling coherent quantum systems.
\end{abstract}

\maketitle

{\em Introduction -- }
Recent experimental advances in the creation of Bose-Einstein 
condensates (BECs) from ultracold atomic gases
have stimulated huge interest in investigating
the coherent many-body dynamics of trapped bosons.
By superposing counter-propagating laser beams, it is
possible to impose extremely well-controlled lattice
potentials on these systems. The precision and flexibility
afforded by this both suggests their use as ``quantum 
matter simulators'' \cite{sims} for systems of interest from other 
areas of physics such as the integer quantum Hall effect \cite{iqhe},
and also permits the clean observation of coherent lattice 
phenomena such as Bloch oscillations \cite{bloch}, 
the formation of repulsively-bound pairs \cite{bound}, and
the Mott transition \cite{mott}.

As well as their purely theoretical interest,
these systems are also highly attractive candidates for applications 
such as quantum information processing due to their long coherence times.
A powerful tool to control their dynamics is provided by
the effect termed ``coherent destruction of tunneling'' \cite{hanggi}, in
which driving the system with an oscillating field has the effect 
of renormalizing the intersite tunneling amplitude. For certain 
parameters of the driving field the tunneling can even be reduced 
to zero, and thus it has been proposed to use this effect to 
control the quantum phase transition between the superfluid and
the Mott state \cite{eckardt,creffield}. 

In this Letter we show how an oscillating driving field can not only
be used to control the {\em amplitude} of the tunneling, 
but also its {\em phase}. This permits the generation 
of a matter-current in analogy to the current induced 
in a conducting ring threaded by a magnetic flux.
The creation of tunneling-phases has been studied before
in BECs using either rotating lattices \cite{rotate}, or
by using the atoms' internal degrees of freedom to
mimic a fictitious magnetic field \cite{internal}.
The scheme we propose is extremely
simple in comparison, requiring only the controlled shaking
of the optical lattice, which has already been demonstrated
in experiment. It also reverses the normal role of 
CDT in a novel way, since the quantum interference effects
which produce CDT are employed here 
to {\em induce} motion, rather than to suppress it.
We study the effect over a range of interaction strengths,
and find it to be present from weak interactions 
right up to the onset of the Mott state.
A surprising feature is that while interactions do reduce the
magnitude of the current, they
do not introduce dephasing or dissipation as seen, 
for example, in Bloch oscillations \cite{dissipation}, but instead 
render the current-generation {\em more} robust. 

{\em Model -- } 
We consider a one-dimensional (1D) optical lattice, in which
the atoms are confined to the lowest Bloch band.
In this case the system can be described very accurately
\cite{jaksch} by the Bose-Hubbard (BH) model
\begin{equation}
\HBH = \sum_{\langle m, n \rangle} \left[ -J \ 
a_m^{\dagger} a_n^{ } + \mbox{H.c.} \right] + 
\frac{U}{2} \sum_{m} n_m (n_m - 1) \ .
\label{hamilton}
\end{equation}
Here $a_m^{ } (a_m^{\dagger})$ are the standard boson 
destruction (creation) operators, $n_m = a_m^{\dagger} a_m^{ }$
is the number operator, and $U$ is the Hubbard-interaction
between a pair of bosons occupying the same site. The tunneling
amplitudes $J$ connect nearest-neighbor sites $\langle m,n \rangle$,
and we take $\hbar =1$.
We now impose a time-dependent potential which
rises linearly across the lattice
\begin{equation}
H(t) = \HBH +  K(t) \sin (\omega t + \theta) \ \sum_{j} j n_j \ ,
\label{drive}
\end{equation}
where $\omega$ is the frequency of the driving
field, and $K(t)$ parameterizes its amplitude. 
Importantly we include the phase of the driving field,
$\theta$, as an additional control parameter.
This form of potential can be produced by
periodically phase-modulating one of the
laser fields providing the optical lattice, 
and has already been used in cold atom 
experiments \cite{morsch} to induce CDT. 

We begin by considering the case of a driving field
of constant amplitude, $K(t) = K$. The Hamiltonian of the system,
Eq. \ref{drive}, is then periodic, with period $T=2 \pi/\omega$.
Accordingly we may use the Floquet theorem 
to write solutions of the time-dependent Schr\"odinger
equation in the form 
$u(t)=\exp(-i \epsilon t) u(t)$, where $u(t)$ is a $T$-periodic
function termed a Floquet state, and $\epsilon$ 
is called a quasienergy. 
In the high-frequency limit, where $\omega \gg (J, U)$ 
is the dominant energy scale of the problem, perturbative
approximations to the Floquet states can be obtained
by solving the Floquet equation for just the driving potential,
and then including $\HBH$ as a perturbation. In this case, 
following the procedure
described in Refs. \cite{creff_pert,sigmund}, 
the Floquet states are given to first-order
by the eigenstates of an 
operator ${\cal H}(t)$ which is identical to $\HBH$, 
but with periodically-varying tunneling amplitudes given by
\begin{equation}
J(t) = J \ \frac{1}{T} \int_0^T 
e^{\pm i K F(\tau,t)} d\tau ,
\label{tunnel}
\end{equation}
where $F(\tau,t) = \int_{t}^{\tau} \sin(\omega t' + \theta) dt'$ 
is the phase accumulated over the interval $(t,\tau)$,
and the $+ / -$ applies to forward/backward hopping.
If we now consider the system stroboscopically, that is,
at discrete times $t = nT$ where $n$ is integer,
the time-dependence of the tunneling amplitudes disappears,
and the system is effectively governed by a 
static Hamiltonian ${\cal H}(0)$, with the Floquet
states $u_j(0)$ playing the role of energy eigenvectors.
Simplifying Eq. \ref{tunnel} then 
reveals that the action of the driving field is to renormalize 
the tunneling amplitudes as
\begin{equation}
\Jeff =  J e^{\pm i (K/\omega) \cos \theta} 
{\cal J}_0(K/\omega) ,
\label{renorm}
\end{equation}
where ${\cal J}_0$ is the zeroth Bessel function of the first kind.
For $\theta = \pi/2$ (cosinusoidal driving) this result
reproduces the familiar Bessel function 
renormalization \cite{eckardt,creffield} 
of the tunneling. An unanticipated result, however,
is that the hopping in general acquires a non-zero {\em phase}, 
which is maximum for $\theta = 0$ (sinusoidal driving).

It may appear suprising that $\theta$ can induce
a non-trivial phase, as changing $\theta$ is 
equivalent merely to shifting the time origin. 
In practice, however, the driving field must be turned
on at a certain time, which thereby does pick out a specific
value for the phase of the driving field.
Since the system is completely coherent, the
effect of this initial condition is not lost during the
subsequent time-evolution, and thus the driving phase $\theta$
can produce different physical results. The central result of our
work is that if $K$ is increased from
zero sufficiently slowly, the Floquet states of the system
are able to adiabatically follow \cite{holthaus_adiabat},
and thereby acquire the $K$-dependent phase
$\phi = K(t) \cos \theta / \omega$. As a result, 
a given initial state can
be transformed into a current-carrying state by slowly ramping
the driving potential from zero to the value that gives
the desired hopping-phase.

\begin{center}
\begin{figure}
\includegraphics[width=0.4\textwidth,clip=true]{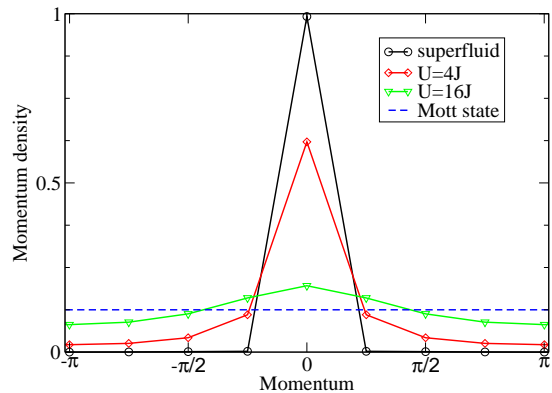}
\caption{Normalized momentum density $\rho(p)/N$ for the
first Brillouin zone of an 8-site lattice holding eight bosons.
For a perfect superfluid the momentum density is sharply peaked, 
showing the presence of long-range coherence. As $U$ is
increased the peak flattens and broadens, until for
$U=16J$ the distribution is almost flat, indicating
the proximity to the formation of a Mott insulator.}
\label{peaks}
\end{figure}
\end{center}

{\em Results -- }
To verify these results, we study the behavior of an $N$-site
BH system by numerically propagating the many-particle wavefunction 
under the time-dependent Hamiltonian (\ref{drive}).
We focus on the case of commensurate filling,
where the number of bosons is equal to $N$, so that in the limit of
large $U$ a well-defined Mott state exists. 
The rapid increase in the dimension of the Hilbert space means
that we could only consider systems of up to $N = 10$, but examining
the results as $N$ is increased reveals that the behaviour we find
is quite insensitive to lattice size.
To probe the behavior of the system, we measure the
single-particle momentum distribution
\begin{equation}
\rho(p,t) = \frac{1}{N} \sum_{m,n}^N e^{i (m-n) p}
\langle \psi(t) | a_m^{\dagger} a_n^{ } | \psi(t) \rangle .
\label{dist}
\end{equation}
This quantity is remarkably size-independent \cite{mom_dist}, 
allowing results from small lattice systems to be reliably 
extrapolated to the thermodynamic limit.
It can be observed directly in experiment
by time-of-flight absorption imaging, and conveniently indicates
whether the system is in the superfluid or Mott-insulator
regime \cite{mott}. When interactions are weak ($U \ll J$), bosons are 
delocalized over the lattice in a superfluid state, 
and the system possesses 
long-range phase coherence. Consequently $\rho(p)$ is sharply 
peaked, as shown in Fig. \ref{peaks}. As the interaction is increased
the peaks broaden and reduce in height, indicating that the
bosons become progressively more localized on the lattice sites.
For sufficiently large values of $U/J$ the atoms localize completely to
form the Mott state, for which 
the momentum distribution is completely flat.
In 1D this phase transition is quite soft,
giving a large range of $U$ over which $\rho(p)$ is peaked.

\begin{center}
\begin{figure}
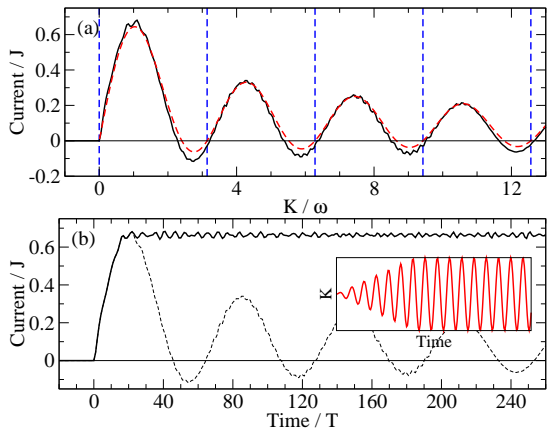

\includegraphics[width=0.4\textwidth,clip=true]{fig2a}
\includegraphics[width=0.4\textwidth,clip=true]{fig2b}
\caption{Current induced by a linearly-ramped driving field
$K(t) = K_0 t$, where $K_0 = 0.05 T^{-1}$,
in an 8-site system with interaction strength $U=8J$.
The driving is sinusoidal ($\theta = 0$).
(a) The induced current (solid black line), shows a 
decaying oscillatory behavior, described well by Eq. \ref{approx_I}
(dashed-red line). As well as at multiples of $\pi$,
marked by the vertical lines, zeros of the current
also occur when ${\cal J}_0(K/\omega)=0$ due to CDT.
(b) As above, the dashed line indicates the current
produced by a continuously ramped field.
Holding $K(t)$ constant after a certain time (shown
schematically in the inset), 
keeps the current at a constant level;
the solid curve shows the effect of ramping the field until
$t=21 T$, which gives the maximum current $I_0$.}
\label{current}
\end{figure}
\end{center}

We begin by considering the case of an intermediate interaction 
strength, $U=8J$. To place the system in the high-frequency
regime we set $\omega = 30 J$, and use a sinusoidal driving
field ($\theta=0$). The precise form in which $K(t)$ is increased
from zero is not important as long as it satisfies the
adiabaticity constraint, and for simplicity we consider  
a linear ramp $K(t)=K_0 t$. 
As noted earlier, we will evaluate all physical quantities
stroboscopically at times $t = nT$.
The system is initialized in its ground-state, and
in Fig. \ref{current}a we plot the expectation value of
the lattice current, 
$I = 2 J \mbox{Im} \langle a_m a_{n}^{\dagger} \rangle$,
as a function of the driving amplitude. For $K(t)=0$ the current
is zero due to the symmetry of the momentum distribution.
As $K$ is increased, however, the peak in $\rho(p)$
is displaced from the center of the Brillouin zone
due to the induced hopping-phase $\phi$,
$\rho(p) \rightarrow \rho(p+\phi)$. As a result $I$ becomes non-zero
due to the imbalance between the left and right-moving
momentum components. 

\begin{center}
\begin{figure}
\includegraphics[width=0.4\textwidth,clip=true]{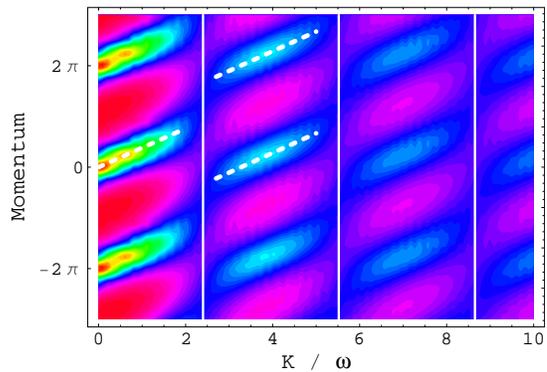}
\caption{Momentum distribution for an 8-site system ($U=8J$)
under sinusoidal driving with a linearly-ramped amplitude.
To guide the eye three Brillouin zones are plotted,
and dashed lines indicate the evolution of the central peak.
The vertical lines mark the zeros of ${\cal J}_0$. 
As $K$ increases the peaks steadily shift
in momentum due to the induced hopping phase, and their amplitude
reduces according to the Bessel function ${\cal J}_0(K/\omega)$.
At $K/\omega=2.4048$ the Bessel function approaches zero and
the system becomes a Mott insulator with a flat 
distribution. Increasing $K$ further causes $\Jeff$ to change
sign, and the peaks reappear with a shift of $\pi$ (see text).
This pattern then repeats.}
\label{peak_shift}
\end{figure}
\end{center}

Following its initial increase,
$I$ displays a damped oscillatory dependence on
$K/\omega$. To interpret this behavior, we show
the corresponding response of the momentum density in
Fig. \ref{peak_shift}. As expected, the initial effect of the ramping
potential is simply to shift the locations of the peaks in the
momentum distribution by inducing the hopping phase.
As the peaks shift, however, their amplitude is reduced
by the Bessel function renormalization of the hopping
amplitude (\ref{renorm}). As $K/\omega \rightarrow 2.4048$, 
the first zero of
${\cal J}_0$, the effective hopping vanishes and the system thus makes
a transition to the Mott state \cite{eckardt} and the momentum density 
flattens. As $K$ is increased further, peaks reappear in $\rho(p)$,
but their location is discretely shifted. This occurs because 
$\Jeff$ has become {\em negative}; writing it
as $\Jeff = |\Jeff| \exp(i \pi)$ clearly indicates that
the peaks in the momentum density will be displaced by $\pi$.
Predicted in Ref.\cite{eckardt_2},
this shift has recently been experimentally
observed for a weakly-interacting ($U \simeq 0.1 J$)
system in Ref.\cite{morsch}. 
The intricate behavior of $I$ in Fig. \ref{current}a thus arises 
from a combination of the shifting location
of the peaks, together with many-particle effects arising from
the competition between $\Jeff$ and $U$.
The roughness visible in the current 
arises from departures from adiabaticity
in the driving. As the ramping rate is decreased,
and so approximates adiabatic evolution more closely, this
roughness is progressively eliminated.
 
For strong interactions, the ground-state of the
system consists approximately of the Mott state, $|11111\dots\rangle$,
with a small admixture of excited states. 
The dominant dynamical processes will be nearest-neighbor 
tunneling between the Mott state and ``particle-hole'' states,
separated from the Mott state by an energy gap of $\sim U$,
where one site is doubly-occupied and one site is empty
(e.g. $|12011\dots\rangle$). Including only these processes,
we can obtain an approximate form for the induced current
\begin{equation}
I \simeq 2 J |\alpha|^2 \ \sin(K \cos \theta/\omega) 
{\cal J}_0(K/\omega) ,
\label{approx_I}
\end{equation}
where $|\alpha|^2$ is the weight of the particle-hole states
in the interacting ground-state. For large $U$,
$|\alpha|^2$ decays as $\sim U^{-1}$ 
as the ground state converges toward the Mott-state.
Using $\alpha$ as a fitting parameter,
we show in Fig. \ref{current}a that this expression indeed
provides an excellent description of the current.
Eq. \ref{approx_I} reveals the two distinct sources for the zeros of 
current; {\cal (i)} when $K/\omega=n\pi$ the
momentum density is symmetrically-peaked at the center of 
the Brillouin zone,
and the positive and negative currents cancel,
{\cal (ii)} when the Bessel function becomes zero, $\Jeff$ is
suppressed and so the tunneling itself is quenched. 

An important consequence of the interaction is that larger values of
$U$ confer increased stability during the ramping process. 
For weak interactions
only extremely slow ramping can be used, or the system will be
excited from its instantaneous ground state and control of the coherent
current will be lost. When $U$ is large, however, the resulting energy
gap isolates the ground-state from the rest of the spectrum
and makes the adiabatic condition easier to attain, thereby
allowing more rapid ramping to be used.

\begin{center}
\begin{figure}
\includegraphics[width=0.4\textwidth,clip=true]{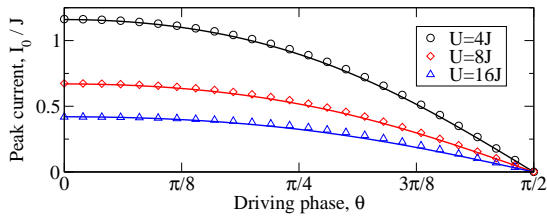}
\caption{The maximum current, $I_0$ induced in an 8-site
system depends on the phase of the driving field, $\theta$.
In all cases $I_0$ is maximized for sinusoidal driving
and is zero for the cosinusoidal case.
Solid lines plot the dependence $\gamma \sin(K \cos \theta /\omega)$,
where $\gamma$ is a fitting parameter,
and show excellent agreement with Eq. \ref{approx_I}.
As $U$ is increased, $I_0$ reduces, 
but even for $U=16J$ the induced current is significant.}
\label{ratchet}
\end{figure}
\end{center}

Differentiating Eq. \ref{approx_I} reveals that the 
maximum current, $I_0$, occurs for
$K/\omega \simeq 1.0311$. In Fig. \ref{current}b 
we show the effect of ramping the value of $K(t)$
up to this value, and then keeping it fixed (see inset),
which maintains the induced current at its final value.
The magnitude of the current depends {\em only} on the final value
of $K/\omega$, and so by regulating this value any desired value of
current within the range $\pm I_0$ can be generated.
In Fig. \ref{ratchet} we show the dependence of $I_0$ on the phase
of the driving $\theta$ for several different values of $U$.
All the curves show the $\sin(K \cos \theta /\omega)$ dependence
expected from Eq. \ref{approx_I}, and 
the magnitude of the current remains significant even 
for large interaction strengths near the onset
of the Mott transition. While smaller values of $U$ allow 
larger currents to be
induced, slower ramping rates must then be used, giving a trade-off
between the two effects in experimental implementations.

{\em Conclusions -- }
We have described a means of inducing a coherent atomic current
by adiabatically-controlling the
renormalization of the intersite tunneling. 
The presence of interactions both stabilizes this mechanism,
and also introduces novel strong-correlation effects. 
To reveal this effect we have employed a stroboscopic
measurement scheme; this implies that in experiment
measurements must be made at well-controlled intervals, and be
sufficiently rapid to reflect the system's instantaneous state.
For typical cold atom systems this would require
temporal control of the order of milliseconds,
which should be easily achievable \cite{morsch}.
If the measurements have insufficient time-resolution,
it would then be appropriate
to consider the time-average of Eq. \ref{tunnel},
which yield the result that
$\langle \Jeff \rangle = J {\cal J}_0^2(K/\omega)$,
and thus the hopping-phase vanishes and the effective hopping
is proportional to the {\em square} of the Bessel function.
For a tilted lattice, it can be shown that
the tunneling is renormalized as ${\cal J}_m^2(K/\omega)$,
where $m \omega$ is the energy difference between
adjacent sites. Interestingly such a dependence
has been recently observed in \cite{tilt} for $m = 1,2$.
While we have focused on the case of bosonic atoms, this mechanism
should be equally applicable to cold fermionic atoms or 
electronic systems, provided that they possess
the required coherence properties.

\bigskip

The authors thank Martin Holthaus for stimulating discussions.
CEC was supported by a Ram\'on y Cajal Fellowship,
and this work was also funded by the MEC (Spain)
through grants FIS2004-05120 and FIS2007-65723.

\end{document}